\documentclass[11pt]{article}
\usepackage{jheppub} 
\usepackage{graphicx} 
\usepackage{amsmath,amsfonts,amssymb,braket,booktabs,empheq,lmodern,mathtools,nccmath,physics,slashed,bm}
\usepackage{wrapfig}
\usepackage{comment}
\usepackage{bbm}
\newcommand*{\df}[1]{\mathrm{#1}}
\newcommand{\beq}{\begin{equation}}
\newcommand{\eeq}{\end{equation}}
\newcommand{\beqn}{\begin{eqnarray}}
\newcommand{\eeqn}{\end{eqnarray}}
\newcommand{\pa}{\partial}
\newcommand{\cB}{\mathcal{B}}
\newcommand{\cK}{\mathcal{K}}
\newcommand{\cS}{\mathcal{S}}
\newcommand{\cN}{\mathcal{N}}

\title{Complexity Growth and the Krylov-Wigner function}

\author[a,1]{Ritam Basu,\note{ritam.basu@tifr.res.in}}
\author[a,2]{ Anirban Ganguly,\note{Visiting student at TIFR, anirban.ganguly@etu.univ-amu.fr}}
\author[a,3]{ Souparna Nath,\note{souparna.nath@tifr.res.in}}
\author[a,4]{ Onkar Parrikar\note{parrikar@theory.tifr.res.in}}

\affiliation[a]{Department of Theoretical Physics, 
Tata Institute of Fundamental Research, 1 Homi Bhabha Road, Mumbai 400005, India}

\abstract{For any state in a $D$-dimensional Hilbert space with a choice of basis, one can define a discrete version of the Wigner function -- a quasi-probability distribution which represents the state on a discrete phase space. The Wigner function can, in general, take on negative values, and the amount of negativity in the Wigner function has an operational meaning as a resource for quantum computation. In this note, we study the growth of Wigner negativity for a generic initial state under time evolution with chaotic Hamiltonians. We introduce the Krylov-Wigner function, i.e., the Wigner function defined with respect to the Krylov basis (with appropriate of phases), and show that this choice of basis minimizes the early time growth of Wigner negativity in the large $D$ limit. We take this as evidence that the Krylov basis (with appropriate phases) is ideally suited for a dual,
semi-classical description of chaotic quantum dynamics at large $D$. We also numerically study the time evolution of the Krylov-Wigner function and its negativity in random matrix theory for an initial pure state. We observe that the negativity rises gradually for a time of $O(D)$ and then saturates close to its upper bound of $\sqrt{D}$.}

\begin{document}
\maketitle
\flushbottom

\parskip=12pt
\section{Introduction}

In general, given a classical phase space, there is a more-or-less standard procedure one can follow to quantize the theory. But given a quantum theory, it is not always clear whether there exists an underlying classical description which is ``useful''. For instance, in the AdS/CFT correspondence, the large $N$ boundary conformal field theory is a quantum mechanical system at strong coupling, so the underlying classical description in terms of classical boundary fields is far from being useful. However, it is a highly non-trivial fact that a new useful classical description emerges in the form of the bulk gravitational theory, with the bulk $\hbar \sim \frac{1}{N^2}$. Given a strongly coupled quantum system, it would be nice to have a general strategy to identify when it admits a dual, semi-classical description, particularly one involving gravity. 

A related question has been studied in quantum information theory -- given a quantum circuit, does it admit a ``useful'' classical description? A quantum circuit consists of some initial state which is sequentially acted upon by a set of quantum gates/operations, followed by a measurement of some observable. The output of the quantum circuit (over several runs) is a probability distribution for the observable of interest. We say that the quantum circuit admits a useful classical description if the same probability distribution can be sampled by an efficient classical algorithm. Here, by efficient we mean that the running time of the classical algorithm should scale polynomially with the spatial size and depth of the quantum circuit. 

A partial answer to this question is given by the Gottesman-Knill theorem \cite{gottesman1998heisenberg, Aaronson:2004xuh} and its generalizations \cite{Mari_2012, Veitch_2012, Pashayan_2015, Wang_2019}. In order to state the result, we need to introduce the notion of the \emph{Wigner function} \cite{Wigner} (see also \cite{Dhar:1992rs, Dhar:1992hr, Lin:2004nb, Balasubramanian:2005mg} for some applications of the Wigner function formalism in holography). The Wigner function $W_{\psi}(q,p)$ for a state $\psi$ is a quasi-probability distribution, which gives a representation of the quantum state in phase space, parametrized here by $(q,p)$. This should be contrasted against the wavefunction $\psi(q)$ of the state, whose absolute square gives a probability distribution over a half-dimensional (Lagrangian) submanifold, parametrized here by $q$. These ideas can also be extended to a quantum system with a finite $D$-dimensional Hilbert space \cite{WOOTTERS19871, Leonard, sphere, quantumcomp, Galois, Gross_2006, classicality}; in this case, one picks an orthonormal basis for the Hilbert space and corresponding to this choice, the Wigner function is defined on a discrete phase space lattice $\mathbb{Z}_D \times \mathbb{Z}_D$.\footnote{The discrete phase space construction works best when $D$ is a prime number, since in this case $\mathbb{Z}_D$ has the structure of a field. The formalism can also be generalized to include powers of primes.} Importantly, the Wigner function is not a true probability distribution because it can take on negative values. Nevertheless, it is a natural quantum analog of a classical probability distribution in phase space; for instance, it is real and unit normalized, and under time evolution follows a quantum generalization of the classical Liouville's equation for the time evolution of phase space densities. When the Wigner function for a state is everywhere non-negative, it acts as a genuine probability distribution, and one might roughly expect such states to be ``classical''. A Wigner function can be associated not only to states, but also to unitary quantum gates (more generally quantum channels) and to measurements. This brings us back to the original question -- when does a quantum circuit admit a useful classical description? It turns out that any ``local'' quantum circuit which uses elements (initial state, quantum gates, measurements etc.) with positive Wigner functions admits an efficient classical implementation \cite{Mari_2012, Veitch_2012, Pashayan_2015, Wang_2019}. Thus, the question of whether one can find an equivalent classical description for the quantum circuit translates to whether one can find a basis in terms of which the elements of a quantum circuit admit positive Wigner function representations. 

On the other hand, given a choice of the basis, classically simulable circuits such as above evolve within a very small subset of states, called \emph{stabilizer states}. Stabilizer states and circuits are not universal -- in order to achieve universality, one needs non-stabilizer elements in the circuit. The negativity of the Wigner function turns out to be an operationally meaningful measure (i.e., a monotone under stabilizer operations) of the non-stabilizer content of a state. Much like entanglement plays the role of a resource in the theory of quantum communication, the negativity of the Wigner function plays the role of a resource in the theory of quantum computation using stabilizer circuits \cite{Veitch_2012, Veitch_2014}. Indeed, Wigner negativity (or ``non-stabilizerness'') is an important feature of various interesting physical systems including conformal field theories \cite{White:2020zoz}, condensed matter systems \cite{Liu:2020yso, Turkeshi:2023lqu, Niroula:2023meg}, topological field theories \cite{Fliss:2020yrd}, and chaotic quantum systems \cite{Leone2021quantumchaosis, OLIVIERO2021127721, Goto}. 

In this note, our aim is to study the time evolution of the negativity of the Wigner function for generic initial states under chaotic Hamiltonian evolution. In this context, we introduce the \emph{Krylov-Wigner function} -- the Wigner function with respect to a specific choice of basis called the Krylov basis \cite{Balasubramanian:2022tpr}.\footnote{See also \cite{Parker:2018yvk, Jian:2020qpp, Rabinovici_2021, caputa2021geometry} for some previous work related to Krylov basis in the context of operator spreading.} Given an initial state $\psi_0$ and a chaotic Hamiltonian $H$, the Krylov basis is generated by repeatedly acting with $H$ on $\psi_0$, 
$$\psi_0,\,H\psi_0,\;H^2 \psi_0 \cdots$$ 
and then orthonormalizing using the Gram-Schmidt process:  
\beqn 
|0_{\cK}\rangle &=& |\psi_0\rangle,\nonumber\\
|1_{\cK}\rangle &=& \frac{1}{\sqrt{N_1}}\left(H|\psi_0\rangle - \langle 0_{\cK}|H|\psi_0\rangle\,|0_{\cK}\rangle\right),\nonumber\\
|2_{\cK}\rangle &=& \frac{1}{\sqrt{N_2}}\left(H^2|\psi_0\rangle - \langle 0_{\cK}|H^2|\psi_0\rangle\,|0_{\cK}\rangle - \langle 1_{\cK}|H^2 |\psi_0\rangle |1_{\cK}\rangle \right),\nonumber\\
 \vdots & & 
 \eeqn
 where $N_i$ are normalization factors. The aspect of chaos that will be important for us is ergodicity, i.e., that the time evolution of a generic initial state should cover the entire (or a large subspace of the) Hilbert space. One of our main results is that the Krylov basis (up to multiplication by individual phases) minimizes the early time growth of the Wigner negativity in the large-$D$ limit. We view this as evidence that the Krylov basis is a potentially good basis for finding an efficient ``close to classical'' description for the quantum time evolution of the initial state.\footnote{However, note that we have not given an algorithm to explicitly demonstrate that the time evolution of the state can be efficiently modeled on a classical computer in this basis.} Our result builds on the work of \cite{Balasubramanian:2022tpr}, where it was shown that the Krylov basis minimizes the amount of early time spreading of the wavefunction under time evolution, a quantity which was referred to as the ``spread complexity''. This already suggests that the wavefunction for the time-evolving state is as peaked as possible in the Krylov basis. One of the motivations of this paper is to show that there is a sense in which the Krylov basis makes the time-evolving wavefunction as classical as possible, and to further explore connections between the work of \cite{Balasubramanian:2022tpr} and computational complexity, say along the lines of \cite{nielsen2005geometric, Nielsen_2006, Jefferson_2017, Chapman_2018, Brown_2018, Balasubramanian:2019wgd, Balasubramanian:2021mxo} (see also \cite{Craps:2023ivc} for recent work in this direction). 
 
 Incidentally, the Krylov basis makes an interesting appearance in a certain model of two dimensional quantum gravity called the double-scaled SYK (DSSYK) model \cite{Cotler:2016fpe, Berkooz:2018jqr, Lin:2022rbf}. The DSSYK model is a model of $N$ Majorana fermions with all to all $p$ body interactions where the couplings are drawn from a Gaussian distribution, and where both $p$ and $N$ are taken to infinity with $\lambda = \frac{2p^2}{N}$ fixed. The model is of interest to quantum gravity because upon averaging over the couplings, the ``chord-diagram'' expansion for the observables in this model, in a certain triple scaling limit, turns into a calculation of observables in Jackiw-Teitelboim (JT) gravity \cite{maldacena2016conformal}. In this correspondence, the Krylov basis in the DSSYK model turns into the ``chord-length'' basis in the dual description \cite{Lin:2022rbf, Rabinovici:2023yex}. In the JT gravity limit, the chord length precisely corresponds to the length of the bulk wormhole, and furthermore, the growth of the spread complexity of \cite{Balasubramanian:2022tpr} corresponds to the growth of the length (what would have been volume in higher dimensions) of the wormhole interior, in line with the ``Complexity = Volume'' conjecture \cite{susskind2014computational, Susskind:2014moa, Stanford:2014jda}. Thus, at least in these models, the dual gravitational description seems to utilise the Krylov basis in a natural way. That the Krylov basis minimizes the early growth of the Wigner negativity in the large-$D$ limit suggests a possibly general underlying principle: \emph{the gravitational variables correspond to a basis which makes the time evolution as classical as possible}. 
 
In order to further explore this proposal for ``gravity as efficient classical implementation of quantum computation'',\footnote{See also \cite{Czech:2017ryf, Caputa:2018kdj} for similar/related ideas.} it would be helpful to study Wigner negativity and its growth in models such as DSSYK or the Saad-Shenker-Stanford matrix model \cite{Saad:2019lba}. A significant obstacle is that the Wigner function and its negativity are difficult to compute analytically, but as a first step towards this goal, here we numerically study the time evolution of the Krylov-Wigner function and its negativity in Gaussian random matrix theory (GUE). We observe that the negativity broadly shows three phases (see figure \ref{fig:neg1}): (i) there is an initial short-lived phase which will not be explored in this paper, (ii) then there is an exponentially long phase where the KW function spreads locally over phase space and its negativity shows a gradual rise, (iii) at a time $t \sim O(D)$, as the Wigner function starts to cover the entire phase space, the negativity starts to saturate and hits a plateau close to its upper bound of $\sqrt{D}$. On the other hand, in a generic basis, the Wigner negativity encounters a sharp ramp at $t=0$ and saturates close to its upper bound value within an $O(1)$ amount of time evolution (see figure \ref{fig:KryCoor}). These results are consistent with the claim that the Krylov basis minimizes the early-time growth of Wigner negativity, and further suggest that this basis is useful for times not scaling exponentially in $\log\,D$.

The plan for the paper is as follows: in section \ref{sec:prelim} we review some background material about discrete Wigner functions and the interpretation of Wigner negativity as a resource for quantum computation. In section \ref{sec:Krylov}, we introduce the Krylov-Wigner function, i.e., the Wigner function defined with respect to the Krylov basis, and show that this choice of computational basis minimizes the early-time growth of the negativity in the large $D$ limit. In section \ref{sec:RMT}, we present some numerical results on the growth of the Krylov-Wigner negativity in random matrix theory. Finally, we end with some discussion on loose ends, open questions and future directions. 

\textbf{Note added in v2}: In v1 of this manuscript, the numerical calculations of the Wigner negativity in random matrix theory displayed a ramp at times of $O(\sqrt{D})$. This was a spurious feature caused by numerical instabilities. In v2, these issues have been fixed and the numerical plots in section \ref{sec:RMT} have been updated. In addition, the discussion below equation \eqref{neg} is improved, and a minor error in equation \eqref{neg2} was fixed. All the main points of the original manuscript remain unaltered. 
 
\section{Preliminaries} \label{sec:prelim}
\subsection{The Wigner function}
Consider a particle moving along on a line under some potential. Traditionally, the classical phase space is the two-dimensional plane coordinatized by the position of the particle $q$ and the momentum $p$. The phase space comes equipped with a symplectic structure, i.e., a closed, non-degenerate 2-form $\omega$, which in the above coordinates is given by
\beq 
\omega = dq\wedge dp.
\eeq 
Equivalently, we can use the above symplectic form to define Poisson brackets on the space of functions on phase space:
\beq 
\left\{f,g\right\}_{PB} = \omega^{IJ} \pa_I f \pa_J g,
\eeq 
where $I=1, 2$ runs over phase space indices and $\omega^{IJ}$ is the inverse matrix corresponding to the 2-form $\omega$ thought of as a matrix. Time evolution in phase space is governed by Hamilton's equations, and correspondingly, given a probability density $u(q,p)$ in phase space, its time evolution is governed by the Louiville equation:
\beq 
\frac{du}{dt} = \left\{H, u\right\}_{PB}.
\eeq 

Quantum mechanically, the particle is described by a Hilbert space. Functions on phase space get promoted to operators on Hilbert space, while the Poisson bracket gets promoted to the commutator. Given a Hamiltonian operator, the time evolution of a density matrix on the Hilbert space is given by
\beq 
\frac{d\rho}{dt} = \frac{i}{\hbar}\left[H, \rho\right]. 
\eeq 
Given a quantum theory, how does one recover the underlying classical phase space? The Wigner function approach is one way to proceed, where one defines a quasi-probability distribution called the Wigner function:
\beq\label{cont_Wig}
W(q,p)=\frac{1}{2\pi\hbar}\int_{-\infty}^{\infty}dy\,\mel{q-\frac{y}{2}}{\rho}{q+\frac{y}{2}}\exp(-\frac{i}{\hbar}py).
\eeq 
Equivalently, one can write the Wigner function as
\beq 
W(q,p) = \frac{1}{2\pi \hbar} \text{Tr}\left(\rho\,A(q,p)\right),
\eeq 
where the operators $A(q,p)$ are called \emph{phase point operators}: 
\beq \label{CPPO}
\langle q'| A(q,p)|q''\rangle = \delta(q-\frac{(q'+q'')}{2}) e^{-\frac{ip}{\hbar}(q'-q'')}.
\eeq 
The phase-point operators are naturally related to the \emph{Heisenberg-Weyl} (HW) displacement operators:
\beq
w(q,p)  = e^{\frac{i}{\hbar}( p\,\mathbf{q} -  q \,\mathbf{p})}= e^{-\frac{i}{2\hbar} pq}Z(p)X(q),
\eeq 
$$Z(p) = e^{\frac{i}{\hbar}p\, \mathbf{q} },\;\;X(q) = e^{-\frac{i}{\hbar}q\, \mathbf{p}}$$
where $\mathbf{q}$ and $\mathbf{p}$ are canonically conjugate operators which satisfy $\left[\mathbf{q},\mathbf{p}\right]=i\hbar$. The phase point operators are symplectic Fourier transforms of the HW operators:
\beq
A(q,p) = \int \frac{dp'dq'}{2\pi \hbar} e^{\frac{i}{\hbar}(p'q - q'p)}w(q',p'),
\eeq 
and can be thought of, heuristically, as being delta-function localized at the phase-space point $(q,p)$.

One can think of the Wigner function as a representation of the quantum state in classical phase space, but note that the Wigner function is, generally speaking, not everywhere positive, so it is not a classical probability distribution. Some useful properties of the Wigner function are:
\begin{enumerate}
    \item The Wigner function is real and unit normalized:
    \begin{equation}
        \int_{-\infty}^{\infty} dx\int_{-\infty}^{\infty} dp\, W(x,p)= 1.
    \end{equation}
    \item Integrating the Wigner function along one-direction gives the probability density. For e.g., 
    \beq 
    \int_{-\infty}^{\infty}dp\,W(x,p) = \langle x|\rho|x\rangle. 
    \eeq 
    \item The time evolution of the Wigner function is controlled by the Moyal equation, which reduces to the Louiville equation in the $\hbar \to 0$ limit:
    \beqn 
    \frac{dW}{dt} &=& \frac{2}{\hbar}\sin\left(\frac{\hbar}{2}(\pa_{x'}\pa_{p}-\pa_{p'}\pa_x)\right)h(x',p')W(x,p) \Big|_{x'=x,p'=p}\nonumber\\
    &=& \left\{h,W\right\}_{PB} + O(\hbar),
    \eeqn
    where $h(x,p)$ is defined as in equation \eqref{cont_Wig}, but with $\rho$ replaced with the Hamiltonian $H$. 
\end{enumerate}

As mentioned above, the Wigner function is not always everywhere positive, and so it is not quite a probability distribution. States for which the Wigner function is positive may be regarded as \emph{classical} states \cite{Anatole}. Hudson's theorem \cite{HUDSON1974249, Soto} shows that the only states for which the Wigner function is everywhere positive are \emph{Gaussian} states:
\beq 
\psi(q) \sim e^{-\frac{1}{2}a q^2 - b q},
\eeq 
where $a, b \in \mathbb{C}$.\footnote{When $a=1$ and $b=0$, this is simply the harmonic-oscillator vacuum, which has a angularly symmetric Wigner function centered at the origin. Turning on an imaginary part for $a$ causes the Wigner function to get squeezed in one direction. The real and imaginary parts of $b$ displace the center of the Wigner function.} Thus, states with everywhere positive Wigner functions form a very restricted class of states in the Hilbert space. But even if we were to start with a classical/Gaussian state, the positivity of the Wigner function will not, in general, be preserved under time evolution. This only happens for at most inhomogenous, quadratic (i.e., free) Hamiltonians \cite{Soto2}. The space of such at most quadratic Hamiltonians generates a group, which we may call the ``Clifford'' group. The Clifford group is the subgroup of the unitary group which normalizes the HW group, or in other words, maps HW operators to HW operators. Any quantum evolution which starts from a Gaussian state and involves only Clifford gates will keep the state within the restricted class of Gaussian states. This leads to a massive reduction in the complexity of the problem: instead of having to solve for the entire wavefunction $\psi(t,q)$ for every $q$ as a function of time, we now only need to keep track of the finite set of parameters $(a(t),b(t))$. In this sense, quantum dynamics gets mapped to generalized classical dynamics for such systems, and time evolution can be efficiently simulated on a classical computer \cite{Bartlett}. 

\subsection{The discrete Wigner function}\label{sec:phasespace}
So far, the discussion has been about infinite dimensional Hilbert spaces, where the phase space is non-compact. Our main focus in this paper will be on quantum systems with finite dimensional Hilbert spaces. For finite dimensional quantum systems, there is a natural analog of the Wigner function, and is referred to as the \emph{discrete} Wigner function \cite{WOOTTERS19871, Leonard, sphere, quantumcomp, Galois, Gross_2006, classicality}. In the finite dimensional case, one defines a \emph{discrete phase space} $\mathcal{P}$ as a lattice of size $D\times D$, where $D$ is the dimension of the Hilbert space. The discrete Wigner function formalism works best when $D$ is a prime number.\footnote{For $D$ either prime or a power of a prime, there exists a field of order $D$. When $D$ is prime, this field is simply $\mathbb{Z}_D$.} For simplicity, we will take $D$ to be a prime number in this work.\footnote{For a Hilbert space with non-prime dimension, we can always embed it in a larger Hilbert space with prime dimension to construct Wigner function representations.} In this case, we think of the discrete phase space $\mathcal{P}$ as $\mathbb{Z}_D \times \mathbb{Z}_D$, with a \emph{symplectic inner product} between two points $\vec{\alpha}=(q,p)$ and $\vec{\beta} = (q',p')$ given by:
\beq 
\left[\vec{\alpha}, \vec{\beta}\right]= pq'-qp'.
\eeq 
Let $\left\{|k\rangle\right\}_{k=0}^{D-1}$ be a particular orthonormal basis for the Hilbert space, which is chosen to begin with. In terms of this basis, one defines a discrete version of the Heisenberg-Weyl (HW) operators:
\beq 
w(q,p) = e^{-\frac{2\pi i}{D}\frac{D+1}{2}qp}Z(p)X(q),
\eeq 
where $(q,p)\in \mathcal{P}$ and 
\beq 
Z(p)|q\rangle = e^{\frac{2\pi i p q}{D}}|q\rangle,\;\;\; X(q) |q'\rangle = |q'+q\rangle.
\eeq
HW operators are closed under multiplication:
\beq \label{eq:multiplication}
w(q,p)w(q',p') = e^{\frac{2\pi i}{D}\frac{(pq'-qp')}{2}}w(q+q',p+p').
\eeq 
In terms of the HW operators, the discrete phase-point operators are defined by a discrete version of the symplectic Fourier transform \cite{WOOTTERS19871}:
\begin{equation}\label{eq:A_Wooters}
A(q,p) = \sum_{k,\ell=0}^{D-1} \widehat{\delta}_{2q,k+l}e^{\frac{2\pi i}{D} (k-\ell)p } \ketbra{k}{\ell}.
\end{equation}
Note that the hatted Kronecker delta $\widehat{\delta}$ is the $\text{mod}\,D$ version, i.e., it is one when $(k+\ell) = 2q\,\text{mod}\,D$, and zero otherwise. We will often use the notation $\vec{\alpha} = (q,p)$ to denote a phase-space point, and correspondingly denote the phase-point operator as $A_{\vec{\alpha}}$. These phase-point operators satisfy the following properties:
\begin{equation}\label{eq:A_properties}
\begin{split}
&\Tr(A_{\vec{\alpha}}) = 1,\\
&\Tr(A_{\vec{\alpha}}A_{\vec{\beta}}) = D\,\delta_{\vec{\alpha},\vec{\beta}},\\
&\frac{1}{D} \sum_{\vec{\alpha}}A_{\vec{\alpha}} = \mathbf{1}.
\end{split}
\end{equation}
The phase-point operators defined above are the finite-dimensional analogs of the phase-point operators in the continuous case (see equation \eqref{CPPO}). In analogy with the continuous case, the discrete Wigner function for a density matrix $\rho$ is now defined in terms of discrete phase point operators as:
\beq 
W_{\rho}(\vec{\alpha}) = \frac{1}{D} \mathrm{Tr}\left(\rho A_{\vec{\alpha}}\right).
\eeq 
Equivalently, any density matrix can be written in terms of its Wigner function as $\rho = \sum_{\vec{\alpha}}W_{\vec{\alpha}}A_{\vec{\alpha}}$. The discrete Wigner function satisfies the following properties:
\begin{enumerate}
\item The discrete Wigner function is real and is unit normalized, i.e.,
\begin{equation}\label{eq:Wigner_norm}
\Tr(\rho) = 1 \implies \sum_{{\vec{\alpha}}} W_{\rho}(\vec{\alpha}) = 1.
\end{equation}
\item Summing over one of the directions, say $p$, reduces the Wigner function to the probability density along the other direction:
\beq 
\sum_{p=0}^{D-1} W_{\rho}(x,p) = \langle x|\rho|x\rangle.
\eeq 
\item The time evolution of the discrete Wigner function follows a discrete version of the Moyal equation:
\beq \label{dis_Moyal}
\frac{dW_{\rho}(\vec{\alpha})}{dt} = \frac{2}{ D}\sum_{\vec{\beta},\vec{\gamma}}\sin\left(\frac{4\pi}{D}\mathcal{A}_{\vec{\alpha}\vec{\beta}\vec{\gamma}}\right)H(\vec{\beta})W_{\rho}(\vec{\gamma}),
\eeq 
where 
\beq 
\mathcal{A}_{\vec{\alpha}\vec{\beta}\vec{\gamma}} = \alpha_2(\gamma_1-\beta_1) + \beta_2(\alpha_1-\gamma_1) + \gamma_2 (\beta_1-\alpha_1),
\eeq 
and the Hamiltonian is expressed in the basis of the phase point operators as $H = \displaystyle\sum_{\vec{\alpha}} H(\vec{\alpha}) A_{\vec{\alpha}}$. To see that the discrete Wigner function satisfies the above differential equation, recall that the density matrix $\rho(t)$ follows the evolution equation (we will henceforth set $\hbar =1$):
\begin{equation}\label{eq:von_Neumann}
\frac{\df{d}\rho}{\df{d}t} = i [\rho,H].
\end{equation}
Substituting the density matrix in terms of its Wigner function in here we get
\beq 
\frac{\df{d}W_{\rho}(\vec{\alpha})}{\df{d}t} = -\frac{i}{ D}\sum_{\vec{\beta},\vec{\gamma}} \mathrm{Tr}\left(A_{\vec{\alpha}} \left[A_{\vec{\beta}},A_{\vec{\gamma}}\right]\right) H(\vec{\beta}) W_{\rho}(\vec{\gamma}).
\end{equation}
This determines the time evolution of the Wigner function in the discrete phase space. Using the explicit form of the phase point operators given in equation \eqref{eq:A_Wooters} above, we then land on equation \eqref{dis_Moyal} \cite{WOOTTERS19871}. 
\end{enumerate}

So far we have discussed Wigner functions for states, but Wigner functions can also be assigned to quantum gates, and more generally to quantum operations/channels \cite{Mari_2012, Wang_2019}. For a quantum channel $\mathcal{E}$, we consider the corresponding Choi-Jamiolkowski state:
\beq 
\sigma_{\mathcal{E}} = (\mathbf{1}\otimes \mathcal{E})(|\Omega\rangle\langle \Omega|),
\eeq 
where 
\beq 
|\Omega\rangle = \sum_{k=1}^D|k\rangle \otimes |k\rangle,
\eeq
is the maximally entangled state on two copies of the Hilbert space. Then, the Wigner function for the quantum operation $\mathcal{E}$ is defined as 
\beq 
W_{\mathcal{E}}(\vec{\beta}|\vec{\alpha}) = \frac{1}{D} \mathrm{Tr}\left(A^T_{\vec{\alpha}}\otimes A_{\vec{\beta}}\,\sigma_{\mathcal{E}}\right) = \frac{1}{D}\mathrm{Tr}\left(A_{\vec{\beta}}\,\mathcal{E}(A_{\vec{\alpha}})\right).
\eeq 
Given an input state $\rho$ with the Wigner function $W_{\rho}(\vec{\alpha})$, the Wigner function of the output state $\mathcal{E}(\rho)$ is given by
\beq 
W_{\mathcal{E}(\rho)}(\vec{\beta}) = \sum_{\vec{\alpha}} W_{\mathcal{E}}(\vec{\beta}|\vec{\alpha}) \,W_{\rho}(\vec{\alpha}).
\eeq 
Similarly, for a positive operator valued measure $\{M_k\}$, the corresponding Wigner function is defined as
\beq 
W_k(\vec{\alpha}) = \mathrm{Tr}\left(M_k\,A_{\vec{\alpha}}\right).
\eeq 
In this way, Wigner functions can be assigned to general elements of a quantum circuit.

\subsection{Negativity of the Wigner function}
As in the continuous case, the discrete Wigner function is also a quasi-probability distribution, in that it can take on negative values. It is natural to regard the set of states with non-negative Wigner functions as being \emph{classical}, but interestingly, the sense in which such states are classical has a computational flavor. Given the choice of an orthonormal basis, states with a non-negative Wigner function are again known to be Gaussian \cite{Gross_2006}:
\beq 
\langle q | \psi\rangle \sim e^{\frac{2\pi i}{D}(aq^2 + bq)},
\eeq 
where $a,b \in \mathbb{Z}_D$. The corresponding discrete Wigner functions are highly constrained. We say that a subspace $\mathcal{L}$ of phase space is an \emph{isotropic subspace} if for any $\vec{\alpha}, \vec{\beta} \in \mathcal{L}$, we have
\beq 
\left[\vec{\alpha},\vec{\beta}\right]=0.
\eeq 
It follows from equation \eqref{eq:multiplication} that the HW group elements corresponding to an isotropic subspace form a subgroup. We say that $\mathcal{L}$ is maximally isotropic if its cardinality is $D$. A maximally isotropic subspace is the discrete analog of a Lagrangian submanifold in phase space. It turns out that positivity is a rather restrictive criterion in the discrete case, and implies that the Wigner function has to be of the form \cite{Gross_2006}:
\beq 
W_{\psi}(q,p) = \begin{cases} & \frac{1}{D} \;\;\cdots\;\;\text{if}\;(q,p)\in \mathcal{L}\\
& 0 \;\;\cdots\;\;\text{if}\;(q,p) \notin \mathcal{L},
\end{cases}
\eeq
where $\mathcal{L}$ is some maximally isotropic subspace. Thus, states with non-negative Wigner functions are ``translationally invariant'' along some ``line'' in phase space. For this reason, such states are called \emph{stabilizer states}, i.e., they are stabilized by the maximally isotropic subgroup of the HW group corresponding to $\mathcal{L}$. 

Note that even though a stabilizer state has a positive Wigner function, in general, unitary evolution will introduce negativity. However, there is a special subgroup of unitaries which preserve the positivity of the Wigner function. This is called the \emph{Clifford group}, and is defined as the subset of unitary operators which map HW operators to HW operators:
\beq 
U w(\vec{\alpha}) U^{\dagger} = w(S(\vec{\alpha})),
\eeq
for some $S:\mathcal{P} \to \mathcal{P}$. In other words, the Clifford group is the normalizer of the HW group. There is a very elegant characterization of the Clifford group in terms of symplectic affine transformations on phase space \cite{Gross_2006}. A \emph{symplectic} transformation $S: \mathcal{P} \to \mathcal{P}$ on phase space is a linear map (with coefficients in $\mathbb{Z}_D$) which preserves the symplectic inner product:
\beq 
\left[S (\vec{\alpha}), S(\vec{\beta})\right] =\left[\vec{\alpha},\vec{\beta}\right]. 
\eeq 
Corresponding to every symplectic map, there exists a unitary operator $\mu(S)$ such that
\beq 
\mu(S)\, w(\vec{\alpha})\, \mu(S)^{\dagger} = w(S(\vec{\alpha})).
\eeq 
Up to a phase, every Clifford unitary is of the form $U = w(\vec{\alpha})\,\mu(S)$. Thus, the Clifford group is essentially a representation of the group of symplectic affine transformations on phase space. The Wigner function transforms covariantly under the Clifford group, namely
\beq  \label{CliffCov}
W_{U\psi}(\vec{\beta}) = W_{\psi}(S(\vec{\beta})+\vec{\alpha}).
\eeq 
Since the Clifford group normalizes the Heisenberg-Weyl group, Clifford rotations preserve the non-negativity of the Wigner function, as can be seen from equation \eqref{CliffCov}. Consequently, Clifford rotations map stabilizer states to stabilizer states. Thus, any quantum circuit which is initialized to a stabilizer state and involves only Clifford operations evolves within a very restricted subset of the Hilbert space. In fact, it is a well-known fact that such quantum circuits can be simulated efficiently on a classical computer. This is the content of the \emph{Gottesman-Knill theorem} \cite{gottesman1998heisenberg, Aaronson:2004xuh}.

More generally, any circuit which involves initializing to a stabilizer state, quantum operations with positive Wigner functions, and measurements defined by positive operator valued measures (POVM) $\{M_k\}$ with positive Wigner functions can be efficiently simulated on a classical computer \cite{Mari_2012, Veitch_2012}. The essential idea is that in a circuit where all elements have positive Wigner functions, the Wigner function of the initial state $\psi_0$ constitutes a genuine probability distribution on phase space, the Wigner function of each quantum operation constitutes a stochastic matrix, and the Wigner function of each POVM $M_k$ constitutes a probability over $k$ for every point in phase space. The outcome of the quantum circuit:
\beq
P_k = \mathrm{Tr}\left(M_k\,\mathcal{E}_m \cdots \mathcal{E}_1(|\psi_0\rangle\langle \psi_0|)\right),
\eeq 
translates to classical stochastic evolution in terms of the Wigner function representations:
\beq 
P_k = \sum_{\vec{\alpha}}\sum_{\vec{\beta_1}\cdots,\vec{\beta}_m}\sum_{\vec{\gamma}}W_{M_k}(\vec{\alpha})\, W_{\mathcal{E}_m}(\vec{\beta_m}|\vec{\beta}_{m-1})\cdots W_{\mathcal{E}_1}(\vec{\beta_1}|\vec{\gamma})\,W_{\psi_0}(\vec{\gamma}).
\eeq
Thus, quantum circuits which use elements with positive Wigner functions essentially reduce to probabilistic classical dynamics. As long as each of these probability distributions can be efficiently sampled,\footnote{This is usually enforced by requiring local quantum operations, i.e., those which act on a few qubits at a time.} then we can also efficiently sample the output probability distribution of the quantum circuit on a classical computer. Interestingly, one can generalize this argument to the case where circuit elements are not Wigner positive, and one finds that the more the Wigner negativity in the circuit, the less efficient the classical simulation \cite{Pashayan_2015, Wang_2019}. 

\subsection{Resource theory of quantum computation}
Based on the above discussion, it is natural to expect that the negativity of the Wigner function is a measure of the non-classicality -- or the inherent quantumness --  of a given state, or of a quantum circuit, and indeed, there is an operational sense in which this is the case. It is clear that circuits which do not involve Wigner negativity only evolve within a very small subset of states in the Hilbert space. Thus, such circuit models cannot be universal. In order to achieve universal quantum computation, one needs an additional resource, namely non-stabilizer states \cite{Bravyi_2005}. A state is said to be \emph{magic} if it is not a stabilizer state. Much like entanglement plays the role of a resource in the theory of quantum communication with local unitary operations and classical communication being treated as ``cheap'', \emph{magic} plays the role of a resource in the theory of quantum computation, with \emph{stabilizer protocols}, i.e., quantum operations using elements with positive Wigner functions (such as stabilizer states, Clifford unitaries and measurements in the computational basis) being treated as being cheap \cite{Veitch_2014}.  

In order to quantify the notion of magic, one defines a \emph{magic monotone} $\mathcal{M}$ as any real-valued measure on density matrices which is, on average, non-increasing under any stabilizer protocol \cite{Veitch_2014}. By on average, we mean that if $\Lambda$ is a stabilizer protocol potentially involving a measurement, and
\beq 
\Lambda(\rho) =  \sigma_i,
\eeq 
with probability $p_i$, then 
\beq 
\mathcal{M}(\rho) \geq \sum_i p_i \mathcal{M}(\sigma_i).
\eeq 
For instance, it could happen that $\mathcal{M}(\rho) < \mathcal{M}(\sigma_i)$; we only require the monotone to decrease on average.\footnote{Note also that the monotone is not necessarily required to be convex.} It turns out that the \emph{logarithmic negativity} of the Wigner function, sometimes also called \emph{mana}:
\begin{equation}
    \label{Mana}
    \mathcal{M}(\psi) = \log\,\cN(\psi),\;\;\;\cN(\psi)= \sum_{x,p} |W_{\psi}(x,p)|,
\end{equation}
is a magic monotone. For e.g., the Wigner negativity remains invariant under a Clifford unitary, and more generally either decreases under either a quantum channel with positive Wigner function, or decreases on average under a POVM with a positive Wigner function \cite{Wang_2019}. The logarithmic negativity has several nice properties. It is evident from the definition that the log-negativity is positive definite, and is zero if and only if the Wigner function is nowhere negative. The log-negativity is also bounded above by $\frac{1}{2}\log \,D$. To see this, we note from equation \eqref{eq:A_properties} that
\begin{equation}
\sum_{\vec{\alpha}} W_{\vec{\alpha}}^2 = \frac{1}{D}\Tr(\rho^2). 
\end{equation}
Together with Jensen's inequality and the fact that $\mathrm{Tr}\,\rho^2 \leq 1$, this gives us the desired bound:
\beq
\sum_{\vec{\alpha}} |W_{\vec{\alpha}}| \leq D \sqrt{\sum_{\vec{\alpha}} W_{\vec{\alpha}}^2} = \sqrt{D} \sqrt{\mathrm{Tr}\,\rho^2}  \leq  \sqrt{D}.
\eeq
Most importantly, the log-negativity has a very direct operational meaning, in terms of \emph{magic distillation}. Any stabilizer protocol which consumes resource states $\rho$ to produce $m$ copies of a target state $\sigma$ with some finite probability, requires, on average, at least $m \frac{\mathcal{M}(\sigma)}{\mathcal{M}(\rho)}$ copies of $\rho$ \cite{Veitch_2014}. This follows from the additivity of logarithmic negativity under tensor factorization, together with the fact that the logarithmic negativity decreases on average under stabilizer operations. Thus, the negativity of the Wigner function constitutes an operationally meaningful notion of the inherent ``quantumness'' of a state. In the rest of this work, we will mainly focus on the negativity as opposed to the log-negativity:
\beq 
\cN(\psi) = \sum_{x,p} |W_{\psi}(x,p)|,
\eeq 
and regard it as a measure of the non-classicality of a state.

\section{The Krylov-Wigner function}\label{sec:Krylov}
Having discussed the importance of Wigner negativity in the theory of quantum computation, our aim now is to study the behavior of the Wigner function and its negativity under time evolution with a chaotic Hamiltonian for generic initial states. But before we go on to study the time evolution of the negativity, we have to make a choice for the basis with respect to which we define the Wigner function.   
\subsection{Krylov basis}
Most physical Hamiltonians come with a natural locality structure, i.e., a tensor product structure for the Hilbert space. For instance, the physical system could be made out of qubits. The Hamiltonian is then written as a linear combination of sufficiently local operators, i.e., products of finitely many single qubit operators. The locality structure, of course, presents a natural candidate for a computational basis. But here, we focus on another potentially natural candidate called the \emph{Krylov basis}, which is defined as follows: given a normalized initial state $\psi_0$, we generate a basis by successively applying the Hamiltonian to $\psi_0$ and then iteratively orthogonalizing using the Graham-Schmidt process. So, we denote $\psi_0$ to be the state $|0\rangle$. Then, we consider the state $H|0\rangle$, orthogonalize it with respect to $|0\rangle$ and normalize to get the state $|1\rangle$:
\beq  \label{krylov1}
|1\rangle = \frac{1}{\sqrt{\langle 0|H^2|0\rangle - (\langle 0|H|0\rangle)^2}}\left(H|0\rangle - \langle 0|H|0\rangle |0\rangle\right),
\eeq 
Then we take the state $H^2 |0\rangle$, orthogonalize with respect to $|0\rangle $ and $|1\rangle$, normalize and call it $|2\rangle$, and keep going. For a chaotic Hamiltonian $H$ without any symmetries, and a generic\footnote{For instance, the initial state should not be an energy eigenstate. Any sufficiently random superposition of energy eigenstates would suffice.} initial state $\psi_0$, we expect that this procedure generates a basis that spans the entire Hilbert space.  

The action of the Hamiltonian in the Krylov basis takes the following simple form:
\begin{equation}\label{KHam1}
H\ket{k} = a_k\ket{k} + b_{k+1}\ket{k+1} + b_{k}\ket{k-1},
\end{equation}
where
\beq 
a_n = \langle n|H|n\rangle,
\eeq 
and the $b_n$ are real, positive coefficients (with the convention that $b_0 = b_D =0$). In matrix form, the Hamiltonian looks like:
\begin{equation}\label{eq:Tridiagonal}
H = \begin{pmatrix}
a_0 & b_1 & 0 & 0 & \cdots & 0 & 0\\
b_1 & a_1 & b_2 & 0 & \cdots & 0 & 0\\
0 & b_2 & a_2 & b_3 & \cdots & 0 & 0\\
0 & 0 & b_3 & a_3 & \cdots & 0 & 0\\
\vdots & \vdots & \vdots & \vdots & \ddots & \vdots & \vdots\\
0 & 0 & 0 & 0 & \cdots & a_{D-2} & b_{D-1}\\
0 & 0 & 0 & 0 & \cdots & b_{D-1} & a_{D-1}\\
\end{pmatrix}
\end{equation}
The Krylov basis has the nice property that it minimizes the ``spreading'' of the wavefunction under time evolution \cite{Balasubramanian:2022tpr}. This can be quantified as follows: consider the time evolved state $|\psi(t)\rangle = e^{-itH}|\psi_0\rangle$. For any ordered, orthonormal basis $\mathcal{B} = \left\{|n_{\cB}\rangle\right\}_{n=0}^{D-1}$, we can defined a measure of the spread of the state $\psi(t)$:
\beq 
C_{\cB}(t) = \sum_{n=0}^{D-1}c_n |\langle n_{\cB}|\psi(t)\rangle|^2,
\eeq 
where $c_n$ is any positive, monotonically increasing weight function; for e.g., a natural choice is to take $c_n=n$. We may think of $C_{\cB}(t)$ as some notion of ``spread complexity'', i.e., a measure of how spread out the state is in the given choice of basis. Let us say we choose $|0\rangle_{\cB} = |\psi_0\rangle$, so that $C_{\cB}(0)$ is set to zero. This does not fix anything about the remaining $(D-1)$ basis vectors of $\cB$. But one could ask for more: one could look for a basis where the early growth of $C_{\cB}(t)$ with respect to time is minimized. In other words, we would like a basis which successively minimizes the time derivatives $\{C'_{\cB}(0), C''_{\cB}(0),\cdots \}$ and so on up to the $(D-1)$th derivative. It was shown in \cite{Balasubramanian:2022tpr} that the Krylov basis minimizes all of these derivatives. In this sense, the Krylov basis quantifies the minimal and inevitable spread in the wavefunction that a state undergoes under quantum evolution. To be clear, adding individual phases to the basis vectors does not change $C_{\cB}(t)$, so any generalized Krylov basis with individual phases also minimizes the spreading. 

Incidentally, the Krylov basis naturally shows up in an interesting way in the context of the double-scaled SYK model \cite{Berkooz:2018jqr, Lin:2022rbf, Rabinovici:2023yex}. The DSSYK model is a model of $N$ Majorana fermions with all-to-all $p$-body interactions. The couplings in the Hamiltonian are chosen randomly from a Gaussian ensemble, and one takes a double-scaling limit where $p \to \infty$, $N\to \infty$ with $\lambda = \frac{2p^2}{N}$ fixed. The model is of particular interest because JT gravity arises in a certain triple scaling limit of the DSSYK model. It turns out that calculations in the DSSYK model can be organized nicely in a particular basis called the ``chord-number'' basis. This basis is the natural generalization of the gravitational length basis (i.e., a basis for the gravitational Hilbert space labelled by the length of the bulk wormhole) in JT gravity. From the perspective of the boundary quantum system, the chord-number basis turns out to be the Krylov basis, and the length of the bulk wormhole maps onto the Krylov index which orders the basis. Thus, the dual ``gravitational'' description of the DSSYK model naturally seems to utilize the Krylov basis. It is natural to wonder what makes the Krylov basis special, i.e., whether there is a deeper underlying reason why the gravitational variables become manifest in the Krylov basis. We will take some baby steps towards addressing this question in section \ref{sec:min}.

\subsection{Krylov phase space}
The results of \cite{Balasubramanian:2022tpr} suggest that the Krylov basis is very natural in the sense that the time-dependent wavefunction is as peaked as possible in this basis. This suggests that the state looks as ``classical'' as possible in the Krylov basis, somewhat like coherent states, except that the latter are peaked in phase space. In order to show that there is a sense in which the Krylov basis makes the time-evolving state as classical as possible, it thus seems natural to study the time-evolution in phase space. Following our discussion in section \ref{sec:phasespace}, we can define a discrete phase space $\mathcal{P} = \mathbb{Z}_D\times \mathbb{Z}_D$ with respect to the Krylov basis, and a corresponding Wigner function for any state $\psi$:
\beq 
W_{\psi}(q,p) = \frac{1}{D}\sum_{k,\ell=0}^{D-1}\widehat{\delta}_{2q,k+\ell}e^{\frac{2\pi i}{D}(k-\ell)p}\langle k|\psi\rangle \langle \psi |\ell\rangle.
\eeq 
Of particular interest to us will be the Wigner function of the time evolved initial state $|\psi(t)\rangle = e^{-itH}|\psi_0\rangle$. We will call this object the \emph{Krylov-Wigner function}. The Krylov-Wigner function gives us a way of visualizing how the state $|\psi\rangle\langle \psi|$ evolves over the discrete Krylov phase space. In words, one can expand the evolved density matrix $|\psi(t)\rangle\langle \psi(t)|$ in terms of the discrete phase-point operators -- which recall form an orthonormal basis for the space of operators -- and the KW-function is the corresponding wavefunction.\footnote{Note that there is a different thing one could have done, which is to construct an operator version of the Krylov basis by constructing the operators $\left\{\rho,[H,\rho],[H,[H,\rho]],\cdots\right\}$ where $\rho = |\psi\rangle\langle\psi|$, and then orthonormalize these with respect to the trace norm. However, this would not give a natural phase-space structure.} It's easy to see that the KW function at time $t=0$ is given by
\beq 
W_{\psi_0}(q,p) = \frac{1}{D}\delta_{q,0}.
\eeq 
In order to see how the Wigner function evolves with time, we can do a Taylor expansion in $t$:
\beqn
W_{\psi(t)}(q,p)&\simeq & \frac{1}{D}\delta_{q,0}- \frac{it}{D}\sum_{k,\ell}\widehat{\delta}_{2q,k+\ell}e^{\frac{2\pi i}{D}(k-\ell)p}\left(\langle k|H|0\rangle\langle 0|\ell\rangle - \langle k|0\rangle \langle 0|H|\ell\rangle\right) )+\cdots\nonumber\\
&=& \frac{1}{D}\delta_{q,0}- \frac{2t}{D}\widehat{\delta}_{2q,1}\sin\left(\frac{2\pi p}{D}\right)+\cdots.
\eeqn
Note that the second term is only non-zero when $q= \frac{D+1}{2}$. So, it would naively seem like the time evolution spreads the Wigner function non-locally. However, this is only a matter of convention and can be easily cured by representing the Wigner function as being defined on the lattice of half-integer separated points $\left\{0,\frac{1}{2},1,\cdots, \frac{D-1}{2}\right\}$, with the understanding that the point $Q=\frac{1}{2}$ in this lattice corresponds to $q=\frac{D+1}{2}$ in $\mathbb{Z}_D$.\footnote{We will use $Q$ to denote half-integer separated points.} This is merely a re-ordering of the phase-point operators, and a convenient one because with respect to this ordering, the spread of the Wigner function has a more local appearance. For instance, at first order in the expansion, the KW function spreads to the point $Q=\frac{1}{2}$, at second order it spreads up to $Q=1$, and so on.  

We can now express the Hamiltonian as a function on the Krylov phase space:
\beqn\label{eq:Hamil_Krylov}
H(Q,p) &=& \frac{1}{D}\Tr(H A(Q,p))\nonumber \\
&=& \frac{1}{D}\sum_{k,\ell=0}^{D-1} \widehat{\delta}_{2Q,k+\ell}e^{\frac{2\pi i}{D} (k-\ell)p }\langle \ell |H|k\rangle \nonumber \\
&=& \frac{1}{D}\sum_{k,\ell=0}^{D-1} \widehat{\delta}_{2Q,k+\ell}e^{\frac{2\pi i}{D} (k-l)p }\Tr\bigg(a_k \delta_{k,\ell} + b_{k+1}\delta_{k+1,\ell} + b_k\delta_{k-1,\ell}\bigg)\nonumber \\
&=& \frac{1}{D}\left(a_Q + b_{Q+\frac{1}{2}}\cos\frac{2\pi p}{D}\right),
\eeqn
where in the third line we have used equation \eqref{KHam1} for how the Hamiltonian acts on the Krylov basis vectors. In phase space, the time evolution of the KW function is controlled by equation \eqref{dis_Moyal}, with the Hamiltonian function given above. 

\subsection{Minimizing the growth of negativity}\label{sec:min}
In this section, we would like to show that there is a sense in which time evolution in the Krylov basis looks as classical as possible. For this purpose, we consider the generalized family of Krylov bases obtained by multiplying individual basis elements by phases:
\beq 
\mathcal{K} = \left\{e^{i\phi_k}|k\rangle \right\}_{k=0}^{D-1},
\eeq 
where $|k\rangle$ is the $k$th basis  vector of \emph{the} Krylov basis, as defined around equation \eqref{krylov1} (with no extra phases). The claim is that for some particular choice of the phases $\{\phi_k\}$, the generalized Krylov basis minimizes the early time growth of Wigner negativity in the large-$D$ limit. This result is quite analogous to the one of \cite{Balasubramanian:2022tpr}, where it was shown that the Krylov basis minimizes the growth of the spread complexity, and our derivation below will follow a similar strategy as in \cite{Balasubramanian:2022tpr}. 

It is worth taking a moment to explain what we mean by ``minimizing the early time growth of the Wigner negativity''. Let $\cB$ be some basis, and let $W_{\cB}$ be the Wigner function defined with respect to that basis. If we wish to minimize the Wigner negativity at $t=0$, then it is clear that we must pick one of the basis vectors to be $\psi_0$, and without loss of generality we call this vector $|0\rangle_{\cB} = |\psi_0\rangle$. With this choice, the Wigner negativity at $t=0$ is minimized. Now, consider the functions:
\beq 
\overline{W}^{(m)}_{\cB}(Q,p) = \sum_{n=0}^m W^{(n)}_{\cB}(Q,p) \frac{t^n}{n!},
\eeq 
where $W^{(n)}_{\cB} = \frac{d^n W_{\cB}}{dt^n}\Big|_{t=0}$. These functions $\overline{W}^{(m)}_{\cB}$ are essentially increasingly accurate Taylor series approximations to the Wigner function in an infinitesimally small neighborhood of $t=0$. Let 
\beq 
\cS_{\cB} = \left\{\cN_{\cB}^{(0)},\cN_{\cB}^{(1)}\cdots, \cN_{\cB}^{(D-1)}\right\},
\eeq 
where
\beq \label{neg}
\cN^{(m)}_{\cB} = \sum_{Q,p} |\overline{W}^{(m)}_{\cB}(Q,p)|.
\eeq 
is the negativity of the $m$th Taylor series approximation. We will show that for any choice of the first basis vector in $\mathcal{B}$, there exists a phase $\phi_1$ such that $\mathcal{N}^{(1)}_\mathcal{K} \leq \mathcal{N}^{(1)}_{\mathcal{B}}$. So, one may as well take $\mathcal{B}$ to agree with $\mathcal{K}$ up to the first basis vector. Similarly, for any choice of the second basis vector of $\mathcal{B}$, there exists a phase $\phi_2$ such that $\mathcal{N}^{(2)}_\mathcal{K} \leq \mathcal{N}^{(2)}_{\mathcal{B}}$, and so on. This is essentially what we mean by the statement that the generalized Krylov basis minimizes Wigner negativity growth at early times.  


Intuitively, if we imagine the time evolution as happening in discrete steps with time intervals $\delta t$, then what we wish to do is to find a class of bases which minimizes the negativity at the first time step, then within that class find a sub-class which minimizes the negativity at the second time-step, and so on. In order to minimize the negativity after the first time step, one would naively want to include the vector $e^{-i\delta t\, H}|\psi_0\rangle$ as a basis element. However, one must first orthogonalize it with respect to $\psi_0$. Similarly, one would naively want to keep including the vectors $e^{-i\,n\,\delta t \,H}|\psi_0\rangle$ as basis vectors at successive time steps, but after first orthogonalizing them with respect to the previous basis vectors. In the $\delta t\to 0$ limit, this is precisely the Krylov basis. The following result makes this intuition precise, but instead of doing it for discrete time steps, we will try and implement this minimization at the level of the above Taylor series approximations in time. Having explained what we mean by minimizing the early time growth of the Wigner negativity, we can now state our result:

\noindent\textbf{Claim}: \emph{In the large-$D$ limit, the (generalized) Krylov basis minimizes the early time growth of Wigner negativity.}

\noindent \textbf{Proof}: Our proof will be inductive in nature. We will assume that for the basis to minimize the first $(d-1)$ elements of $\cS_{\cB}$, the basis vectors $\{|n_{\cB}\rangle\}_{n=0}^{d-1}$ must be of the generalized Krylov form, with some particular choice of phases. Now our aim is to show that the requirement that the function $\cN^{(d)}_{\cB}$ be minimized forces us to pick the $d$th basis vector to be the $d$th Krylov basis vector, up to a phase. To this end, consider a basis $\mathcal{B}$ such that its first $d$ basis vectors agree with that of the Krylov basis $\mathcal{K}$, up to phases which are determined by the requirement that the first $(d-1)$ elements of $\cS_\cB$ are minimized. The time-derivatives of the Wigner function at $t=0$, with respect to the basis $\cB$, are given by:
\beq 
W_{\mathcal{B}}^{(m)}(Q,p) = \frac{1}{D}\sum_{k,\ell=0}^{D-1}\widehat{\delta}_{2Q,k+\ell}e^{\frac{2\pi i}{D}(k-\ell)p}\sum_{n=0}^m (-1)^n i^m \frac{m!}{n!(m-n)!}\langle \ell_{\cB}| H^n |\psi(0)\rangle \langle \psi(0)| H^{m-n}|k_{\cB}\rangle. 
\eeq 
For $m \leq (d-1)$, the derivatives of the Wigner function for $\mathcal{B}$ agree precisely with that of $\mathcal{K}$. This means that the Taylor expansion of the Wigner function around $t=0$ agrees up to and including $O(t^{d-1})$ terms. The first deviation happens at $m=d$:
\beq 
W_{\mathcal{B}}^{(d)}(Q,p) = \frac{1}{D}\sum_{k,\ell=0}^{D-1}\widehat{\delta}_{2Q,k+\ell}e^{\frac{2\pi i}{D}(k-\ell)p}\sum_{n=0}^d (-1)^n i^d \frac{d!}{n!(d-n)!}\langle \ell_\cB| H^n |\psi(0)\rangle \langle \psi(0)| H^{d-n}|k_\cB\rangle .
\eeq 
In particular, the terms which are different from the generalized Krylov basis correspond to $n=0$ and $n=d$. Now,
\beq 
H^d |\psi(0)\rangle = c_d |d_{\cK}\rangle + |\chi\rangle,
\eeq 
where $|\chi\rangle$ is some linear combination of the Krylov basis vectors of order lower than $d$, and
$$ c_d = e^{-i\phi_d}\prod_{n=1}^{d} b_n,$$
where the phase $\phi_d$ is the phase of the $d$th basis vector in $\cK$ which is arbitrary for now, but will be fixed momentarily. Thus, we get
\beqn 
W_{\mathcal{B}}^{(d)}(Q,p) &=& \frac{1}{D}\sum_{k,\ell=0}^{D-1}\widehat{\delta}_{2Q,k+\ell}e^{\frac{2\pi i}{D}(k-\ell)p}\sum_{n=1}^{d-1} (-1)^n i^d \frac{d!}{n!(d-n)!}\langle \ell_{\cK}| H^n |\psi(0)\rangle \langle \psi(0)| H^{d-n}|k_{\cK}\rangle \nonumber\\
&+& \frac{1}{D}\sum_{\ell=0}^{D-1}\widehat{\delta}_{2Q,\ell}\left(e^{-\frac{2\pi i}{D}\ell p}  (-i)^d  \langle \ell_{\cK}| \chi\rangle + e^{\frac{2\pi i}{D}\ell p}  i^d\langle \chi|\ell_{\cK}\rangle\right)\nonumber\\
&+& \frac{i^d}{D}\sum_{k=0}^{D-1}\widehat{\delta}_{2Q,k}\left(e^{\frac{2\pi i}{D}kp}   c_d^*\langle d_{\mathcal{K}}|k_{\mathcal{B}}\rangle + e^{-\frac{2\pi i}{D}kp}  (-1)^d c_d \langle k_{\mathcal{B}}| d_{\mathcal{K}}\rangle\right).
\eeqn
We can re-write this as
\beq \label{dWig}
W_{\cB}^{(d)}(Q,p) = W_{\cK}^{(d)}(Q,p)+ \frac{i^d}{D}\sum_{k=d}^{D-1}\widehat{\delta}_{2Q,k}\left(c_d^*e^{\frac{2\pi i}{D}kp}   (\langle d_{\mathcal{K}}|k_{\mathcal{B}}\rangle-\delta_{k,d}) + c_d e^{-\frac{2\pi i}{D}kp}  (-1)^d  (\langle k_{\mathcal{B}}| d_{\mathcal{K}}\rangle-\delta_{k,d})\right).
\eeq 
Note that the sum above is restricted to $k\geq d$ because for $k<d$, the basis vectors in $\cB$ are proportional to the Krylov basis vectors, and thus orthogonal to $|d_{\cK}\rangle$. Let us temporarily fix $\phi_d$ by requiring that $\langle d_{\cB}| d_{\cK}\rangle$ be a real number in the range $[0,1]$. This just picks out a particular member in the class of generalized Krylov bases. Our aim is to show that this basis already has lower negativity than $\cB$, and then minimizing over $\phi_d$ will only lower the negativity. Let us  denote:
\beq 
\langle k_{\cB}|d_{\cK}\rangle =\delta_{d,k}+ \alpha_k,\;\;\cdots\;\;(k\geq d)
\eeq 
where $\alpha_d \in \mathbb{R}$ and $-1\leq \alpha_d \leq 0$ (be choice of $\phi_d$), and $$(1+\alpha_d)^2+\sum_{k=d+1}^{D-1} |\alpha_k|^2 =1. $$
 We can thus write equation \eqref{dWig} as follows:
\beq \label{dDer1}
W_{\cB}^{(d)}(Q,p) = \begin{cases} & \displaystyle W_{\cK}^{(d)}(Q,p)+ \frac{2(-1)^{\frac{d}{2}}}{D}|c_d||\alpha_{2Q}|\Theta_{2Q \geq d}\cos\left(\frac{2\pi (2Q)p}{D} + \theta_{2Q} +\phi_d\right) \;\; \cdots (d\;\text{even})   \\
& \displaystyle W_{\cK}^{(d)}(Q,p)+ \frac{2(-1)^{\frac{d-1}{2}}}{D}|c_d||\alpha_{2Q}|\Theta_{2Q \geq d}\sin\left(\frac{2\pi (2Q)p}{D} + \theta_{2Q}+\phi_d\right)\;\;\cdots (d\;\text{odd}),
\end{cases}
\eeq 
where $\alpha_k = |\alpha_k|e^{i\theta_k}$, $c_d=|c_d|e^{-i\phi_d}$ and $\Theta_{2Q \geq d}$ is a step function which is one only when $2Q \geq d$ and zero otherwise. The negativity at this order is given by
\beq 
\mathcal{N}^{(d)} = \sum_{Q,p}|\overline{W}^{(d)}_{\cB}(Q,p)|.
\eeq 
Including terms up to $O(t^{d-1})$, the Wigner function has non-trivial support till $2Q=d-1$. After including the $O(t^d)$ terms, the Wigner function in this region is exactly equal to that of the Wigner function in the generalized Krylov basis, i.e., the difference in basis vectors beyond the $(d-1)$the vector does not manifest in this region:
\beq 
\overline{W}^{(d)}_{\cB}(Q,p) = \overline{W}^{(d)}_{\cK}(Q,p),\;\;\; \forall \;\;2Q < d.
\eeq 
So we need only look at the negativity on the region $2Q \geq d$. In this region, there is no contribution from lower order derivatives, so we wish to compare $\sum_{p,2Q \geq d}|W^{(d)}_{\cB}(Q,p)|$ with $\sum_{p,2Q \geq d}|W^{(d)}_{\cK}(Q,p)|$. 

Let us consider $d$ even; the $d$ odd case works analogously. From the triangle inequality applied to equation \eqref{dDer1}, we get
\beq 
\left|W_{\cK}^{(d)}(d/2,p)\right| \leq \left|W_{\cB}^{(d)}(d/2,p)\right| + \frac{2}{D}|c_d| |\alpha_d| \left|\cos\left(\frac{2\pi dp}{D} + \theta_d + \phi_d\right)\right|. 
\eeq 
Thus, we find
\beqn \label{neg2}
\sum_{Q\geq d/2}|W_{\cK}^{(d)}(Q,p)| &\leq &  \sum_{Q\geq d/2}|W_{\cB}^{(d)}(x,p)| + \frac{2}{D}|c_d|\Big\{ |\alpha_d| |\cos\left(\frac{2\pi dp}{D} + \theta_d + \phi_d\right)| \nonumber\\
&-&  \sum_{Q > d/2}|\alpha_{2Q}| |\cos\left(\frac{4\pi Qp}{D} + \theta_{2Q}+\phi_{d}\right)|\Big\}.
\eeqn
Now we need to sum the above inequality over the momentum direction. In the large $D$ limit, the sums over the terms in the curly brackets can be approximated by integrals, and this gives
\beq
\sum_{p, Q\geq d/2}|W_{\cK}^{(d)}(Q,p)| \leq  \sum_{p,Q\geq d/2}|W_{\cB}^{(d)}(Q,p)| + \frac{4|c_d|}{\pi} \left( |\alpha_d| - \sum_{k > d}|\alpha_{k}| \right).
\eeq
Our goal now is to show that $|\alpha_d| \leq \sum_{k>d}  |\alpha_k|$. We know that
\beq 
(1+\alpha_d)^2 + \sum_{k> d} |\alpha_k|^2 = 1, 
\eeq 
so using the fact that $\alpha_d$ is real and negative, we get
\beq 
-2|\alpha_d| + |\alpha_d|^2 + \sum_{k> d} |\alpha_k|^2 = 0.
\eeq 
Since $|\alpha_d|$ and $|\alpha_k|$ are all positive numbers less than 1, we get
\beq 
-|\alpha_d|  + \sum_{k> d} |\alpha_k|  \geq 0,
\eeq 
which then implies $|\alpha_d| \leq \sum_{k> d} |\alpha_k|$. Thus, we have shown that for any given basis $\cB$, we can choose the phase of the $d$th generalized Krylov basis vector in a way such that we can always make the negativity of the generalized Krylov basis smaller than the given basis $\cB$. Thus, if we pick the phase of the $d$th Krylov basis vector by simply minimizing the negativity over the choice of this phase, we are guaranteed that there exists a choice of a generalized Krylov basis which minimizes the negativity up to this order.\hspace{\fill}$\Box$

In summary, the generalized Krylov basis with appropriately chosen phases minimizes the early growth of the Wigner negativity in the large-$D$ limit. So, if one's goal is to try as hard as possible to simulate quantum dynamics on a classical computer, then it is natural to expect that the Krylov basis (with appropriately chosen phases) would be a good bet for the computational basis. An important caveat to bear in mind is that we have not given an algorithm here to demonstrate efficient classical simulation of time evolution in the Krylov basis, although we take our results as some evidence in this direction. Still, it is natural to wonder whether this is what makes the Krylov basis special, and in particular, a natural candidate for a ``dual gravitational basis'' for strongly coupled quantum systems, with equation \eqref{eq:Hamil_Krylov} being the natural analog of the gravitational Hamiltonian. Remarkably, in the DSSYK model, this is exactly what happens -- the dual ``gravitational'' description consists of the chord-length basis, which precisely coincides with the Krylov basis \cite{Lin:2022rbf, Rabinovici:2023yex}. So, it seems plausible that \emph{the gravitational variables correspond to a choice of basis which makes classical simulation of quantum dynamics as efficient as possible}. There are many important details which need to be worked out to make this proposal more concrete, some of which are listed in the discussion section.  

\section{Krylov-Wigner negativity in random matrix theory}\label{sec:RMT}
In this section, we will study the growth of the negativity of the KW function in random matrix theory. The DSSYK model or the SSS matrix model would perhaps be the most natural models to consider. But as a first step, here we will consider Gaussian random matrix theory. In the following sections, we will consider the Krylov basis without any added phases. It would be interesting to explore addition of phases to these results. 
\subsection{Setup}
We start with some working basis, which we will call the ``coordinate basis''. We pick the initial state $\psi_0$ to coincide with one of the states in this basis, say, $|0\rangle$. Next, we pick a $D\times D$ matrix from the Gaussian Unitary ensemble (GUE) and declare that as the Hamiltonian in the coordinate basis. The GUE corresponds to the following probability distribution over the space of Hermitian matrices:
\begin{equation*}
 p(H)=   \frac{1}{Z_\mathrm{GUE}} e^{-\frac{D}{2}\mathrm{Tr}(H^2)}
\end{equation*}
where $Z_\mathrm{GUE} = 2^{\frac{D}{2}}(\frac{\pi}{D})^{\frac{D^2}{2}}$ is the normalization. Since the Hamiltonian is random, we expect that the $|0\rangle$ state is a sufficiently generic superposition of energy eigenstates of $H$. In the context of holography, we can think of this choice of the initial state as a typical black hole microstate at infinite temperature, along the lines of \cite{Kourkoulou:2017zaj}. We now build the Krylov basis with $|0\rangle$ as the first state, and then successively acting with the Hamiltonian and orthonormalizing using the Gram-Schmidt process. Having built a Krylov basis for the Hilbert space, we then construct the KW function and compute its negativity as a function of time:
\beq 
\mathcal{N}(t) = \sum_{Q,p} |W_{\psi(t)}(Q,p)|,
\eeq
where $|\psi(t)\rangle = e^{-itH}|\psi_0\rangle$.

\begin{figure}
    \centering
\begin{tabular}{c c}
    \includegraphics[height=5.6cm,width=0.5\linewidth]{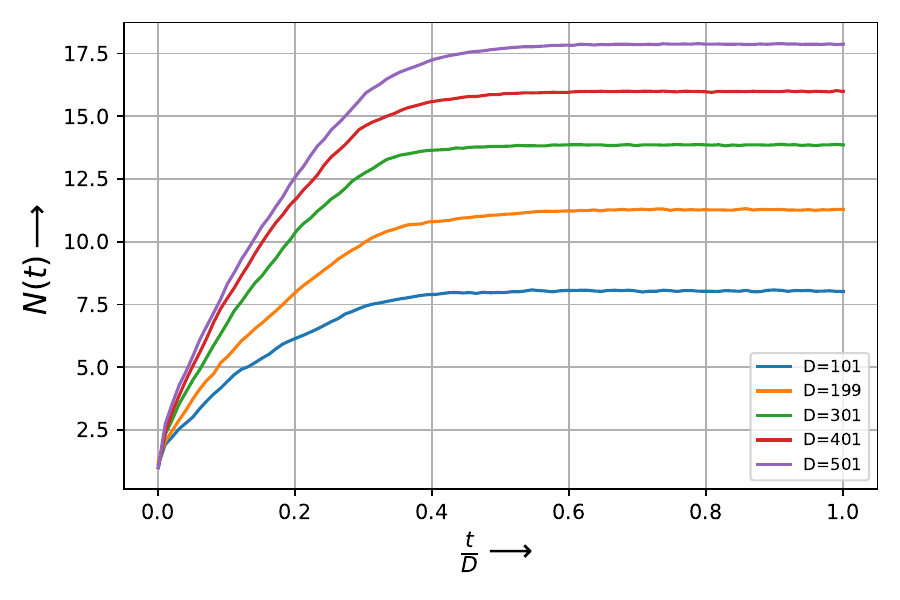} & \includegraphics[height=5.6cm,width=0.5\linewidth]{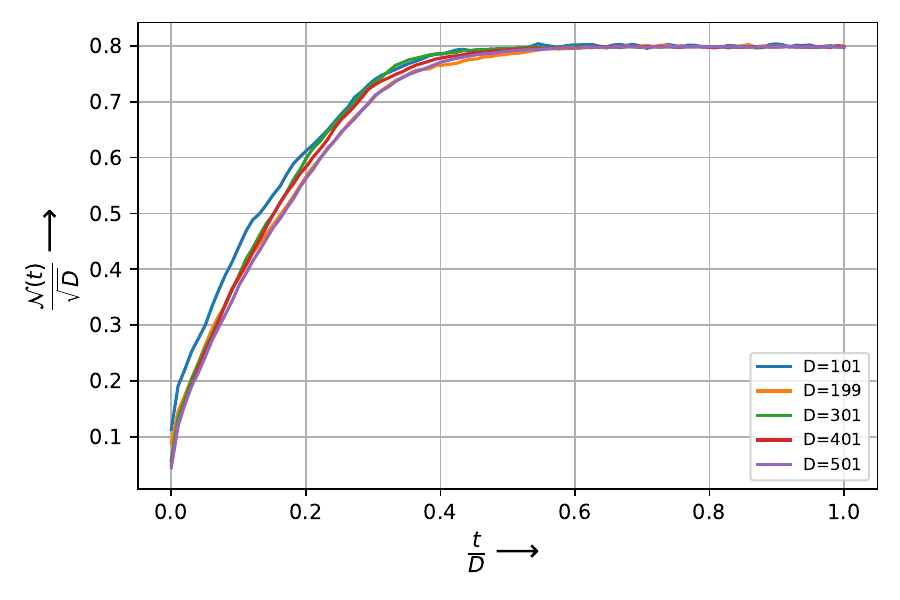}
    \end{tabular}
    \caption{Negativity of the Wigner function $\mathcal{N}$ as a function of the re-scaled time $\frac{t}{D}$ for a randomly chosen Hamiltonian from the GUE for various values of $D$.}
    \label{fig:neg1}
\end{figure}


\subsection{Negativity growth}
In figure \ref{fig:neg1}, we plot the negativity of the Wigner function as a function of the re-scaled time $\frac{t}{D}$. The numerical plots broadly show at least three distinct phases for the negativity: (i) a small early time phase which will not be explored in this paper, (ii) an exponentially long phase phase where the negativity grows gradually with time, and (iii) saturation to a plateau at approximately $\cN_{\text{sat.}}\sim 0.8\sqrt{D}$ at late times ($t\sim O(D)$). Note that the saturation value is close to but less than the upper bound of $\sqrt{D}$.
\begin{figure}
    \centering     \includegraphics[height=15cm]{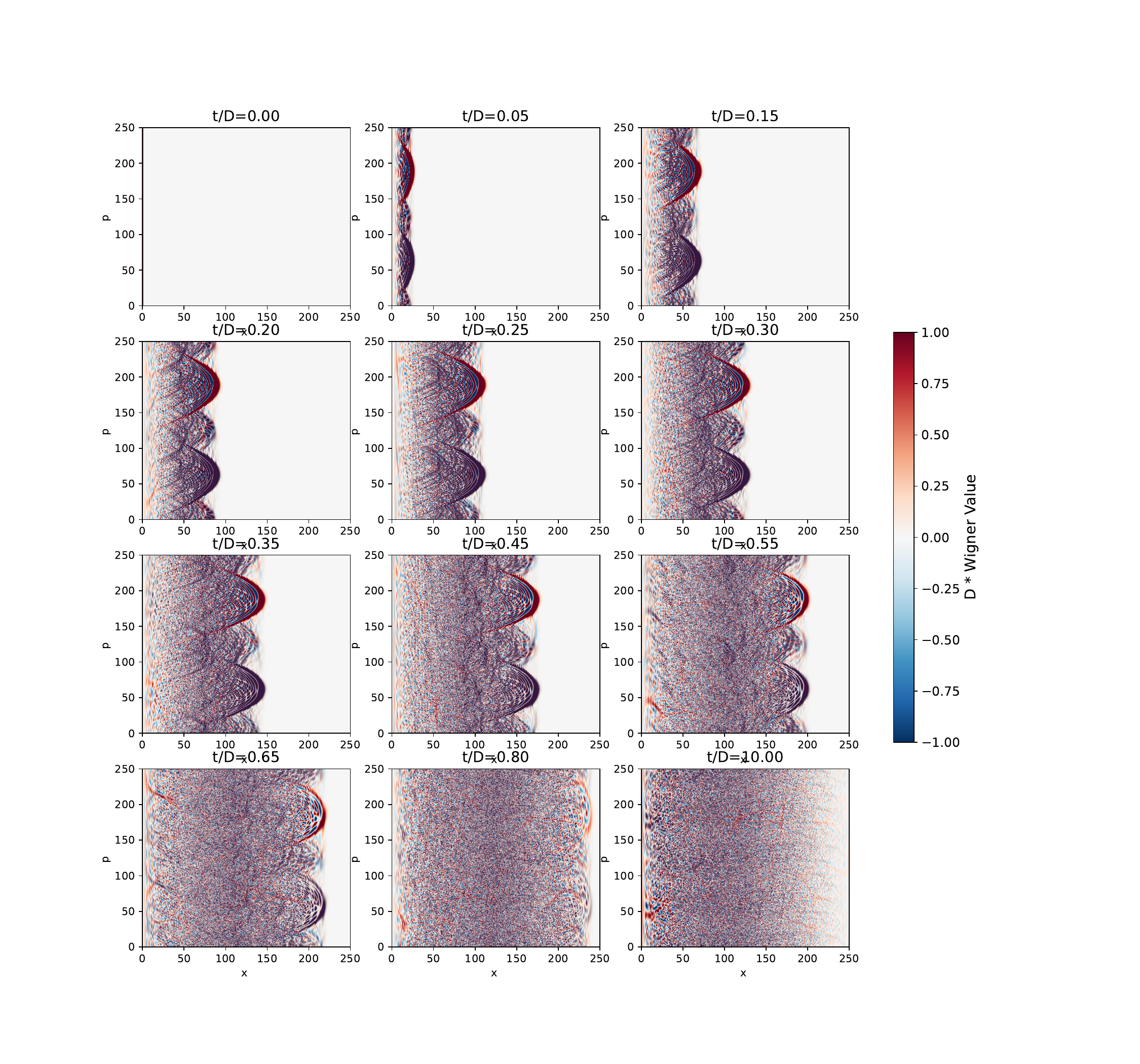}
    \caption{Density plots of $D$ times the Krylov-Wigner function in phase space for $D=251$ for various values of time corresponding to $\frac{t}{D} = (0.0,0.05,0.15,0.2,0.25,0.3,0.35, 0.45,0.55,0.65,0.8,10)$.
    }
    \label{fig:spreading}
    \end{figure}

\begin{figure}
    \centering    
    \begin{tabular}{c c}
    \includegraphics[height=5.6cm,width=0.5\linewidth]{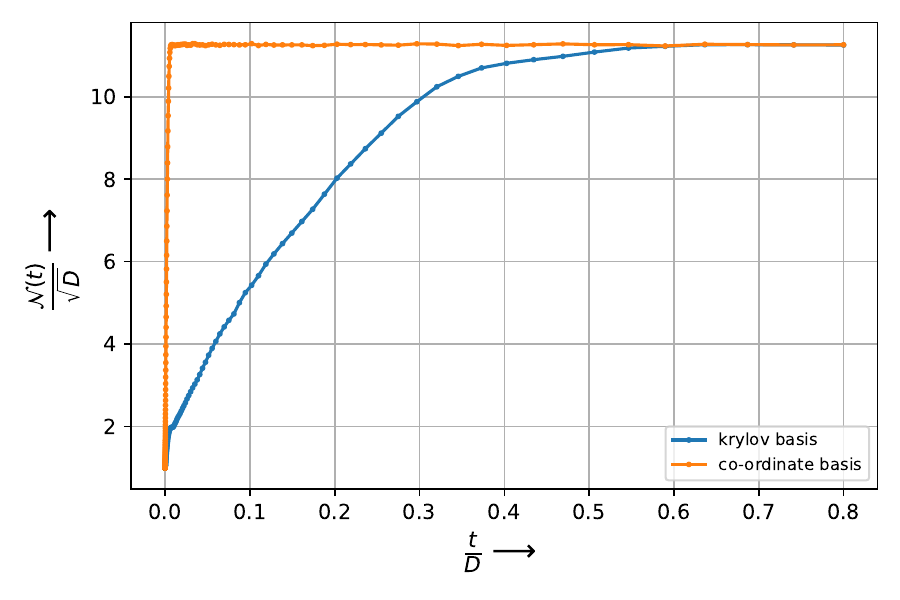} &  \includegraphics[height=5.6cm,width=0.5\linewidth]{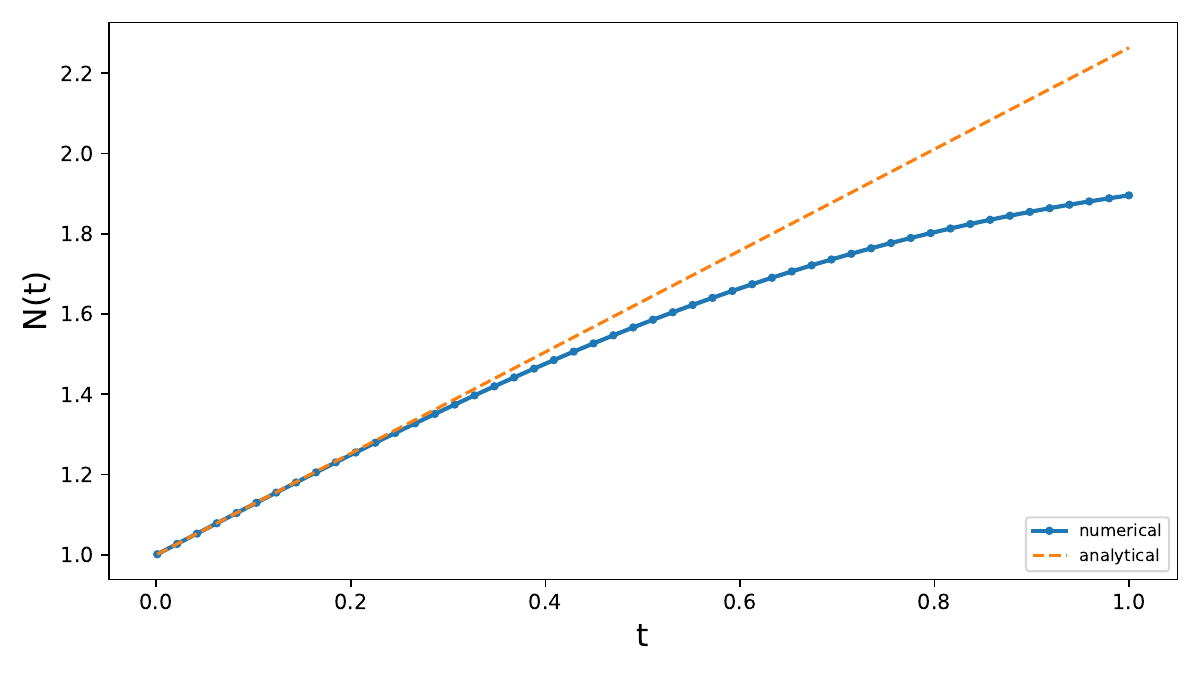} 
    \end{tabular}
    \caption{(Left) The negativity of the Wigner function with respect to the coordinate basis in comparison with that of the Krylov-Wigner function for $D=199$. (Right) The KW negativity for small times and $D=199$. The blue curve is the numerical result, while the dashed orange curve is a straight line with slope $\frac{4}{\pi}$.}
    \label{fig:KryCoor}
\end{figure}

At $t=0$, the Wigner function is localized at $Q=0$, but spread uniformly over the momentum direction:
\beq 
W(0,Q,p) = \frac{1}{D} \delta_{Q,0}.
\eeq 
As time increases, the Wigner function spreads in phase space (see figure \ref{fig:spreading}). In the spreading phase, the Wigner negativity grows gradually with time. This growth seems to be much slower than would be the case had we picked any other computational basis. For instance, in figure \ref{fig:KryCoor} we compare the growth of the negativity between the Krylov basis and the original coordinate basis, i.e. the basis with respect to which the entries of the Hamiltonian are picked to be random. Note that the negativity in the coordinate basis rises sharply starting at $t=0$, and saturates to the same exponentially large value of about $0.8 \sqrt{D}$ within a time of $O(1)$. We expect this to be the behaviour in any generic basis.

In contrast, the KW negativity shows a much more gradual rise, with an $O(1)$ slope at $t=0$. To compute the slope, recall that the Wigner function can be expanded perturbatively in time as follows:
\beqn
W(t,Q,p) &=&  \frac{1}{D}\sum_{k,\ell} \widehat{\delta}_{2Q,k+\ell} e^{\frac{2\pi i}{D} (k-\ell)p } \left(\bra{\ell}\rho(0)\ket{k} -it\bra{\ell}[H,\rho(0)]\ket{k} +\mathcal{O}(t^2)\right)\\
&=& \frac{1}{D}\sum_{k,\ell} \widehat{\delta}_{2Q,k+\ell} e^{\frac{2\pi i}{D} (k-\ell)p } \Big(\langle \ell|0\rangle \langle 0| k\rangle -it(\langle \ell| H |0\rangle \langle 0|k\rangle -\langle \ell| 0\rangle \langle 0|H|k\rangle) +\mathcal{O}(t^2)\Big).\nonumber
\eeqn
Using 
\beq 
H|0\rangle = a_0 |0\rangle + b_1|1\rangle,
\eeq 
we get
\beqn
W(t,Q,p) &=& \frac{1}{D}\delta_{2Q,0} -\frac{2tb_1}{D} \sin\left(\frac{2\pi p}{D}\right)\delta_{2Q,1}+ O(t^2).
\eeqn
At linear order in time, the Wigner function is only supported at $Q=0$ and $Q=\frac{1}{2}$, so we can compute the negativity as follows:
\begin{equation}
\cN = 1 + \frac{2t b_1}{D} \sum_{p=0}^{D-1} \left|\sin(\frac{2\pi p}{D})\right|+O(t^2).
\end{equation}
In the large $D$ limit, we can convert the sum into an integral, and we get
\beq
\cN = 1 + \frac{2t}{2\pi} b_1 \int_{0}^{2\pi} d\theta|\sin(\theta)| = 1 + \frac{4t}{\pi}b_1+O(t^2).
\eeq
Thus, the negativity always starts rising linearly with a slope proportional to $b_1$. We expect the expectation value of $\overline{b_1}$ in GUE to be $O(1)$ \cite{Balasubramanian:2022tpr, Balasubramanian:2022dnj}, so that the slope is approximately equal to $\frac{4}{\pi}$ times an $O(1)$ constant (see figure \ref{fig:KryCoor} for comparison with numerics). We can also give a non-linear approximation for the early time behaviour of the KW function if we assume all the $b_n$ for small enough $n$ are close to one (see Appendix \ref{sec:app}). 

As $\frac{t}{D}$ becomes $O(1)$, the Wigner function starts spreading out over the entire phase space. Up until this point, the negativity is smaller than the upper bound of $\sqrt{D}$. Once the spreading wave covers all of phase space, the negativity starts to become $O(\sqrt{D})$ and saturates to a value of approximately $0.8\sqrt{D}$, close to its maximum possible value of $\sqrt{D}$. In this plateau phase, the Wigner function has spread over the entire phase space uniformly, with a large number of positive and negative pockets uniformly distributed throughout. Our numerical results are consistent with the claim that the Krylov basis minimizes the early-time growth of Wigner negativity, and further suggest that this basis is useful for times not scaling exponentially in $\log\,D$. Note that the growth of the Wigner negativity for an exponential (in $\log\,D$) amount of time followed by saturation to an exponentially large value at exponential time follows the generally expected behavior of measures of computational complexity \cite{Brown_2018}.

\section{Discussion}


    \label{fig:}


In this work, we have introduced and studied the Krylov-Wigner function, i.e., the discrete Wigner function defined with respect to the Krylov basis. We showed that with an appropriate choice of phases, the Krylov basis minimizes the growth of the negativity of the Wigner function for large $D$ quantum systems. We then studied the negativity of the Krylov-Wigner function in Gaussian random matrix theory and observed broadly three phases in its time evolution. Much like circuit complexity \cite{Brown_2018, Balasubramanian:2019wgd, Balasubramanian:2021mxo}, the KW negativity also keeps growing till a time exponential in $\log\,D$, and then saturates at an exponentially large value. We end with some remarks on the outlook and future directions.

\subsection{Gravity as efficient computation}

The Krylov basis has appeared recently in an interesting way in the context of the DSSYK model \cite{Berkooz:2018jqr}, where it emerges naturally in the dual gravitational description as the eigenstates of the length operator of the bulk wormhole \cite{Lin:2022rbf, Rabinovici:2023yex}. We hope that our results shed new light on this point. The negativity of the Wigner function is a measure of the intrinsic non-classicality in a state. The fact that the Krylov basis minimizes the early-time growth of the Wigner negativity in the large-$D$ limit suggests that this basis is ideally suited for a dual, semi-classical description of chaotic quantum dynamics at large $D$. Here, the aspect of chaos that is important is ergodicity, i.e., we want the Krylov basis to span the entire (or at least a sufficiently large subspace of the) Hilbert space. From this perspective, it is natural to think of the Krylov phase space and the corresponding Hamiltonian as the dual ``gravitational'' theory, which seeks to implement the boundary quantum dynamics on a classical computer to the best extent possible, supplemented with some additional non-classical resource. This motivates us to propose a new computational principle for the emergence of spacetime and gravity -- \emph{the gravitational variables correspond to a choice of basis which makes classical implementation of quantum computation as efficient as possible}. Admittedly, at this stage, this a rather vague and rudimentary proposal, and there are several issues which need to be addressed in order to make it more concrete:
\begin{enumerate}

\item Firstly, this ``dual description'' in the Krylov basis appears to come with a discrete phase space at finite $D$. Presumably, continuum gravity emerges from an appropriate coarse-graining in the large-$D$ limit. Roughly, one would like to first send $D\to \infty$. If the Krylov coefficients $b_n$ (for $n$ not scaling with $D$) are sufficiently smooth and slowly varying functions of $n$ (perhaps after ensemble averaging) with some effective Planck scale $\lambda$ which cuts off the large-$n$ growth of these coefficients, then one can take a continuum limit where $\lambda \to 0,\, n\to \infty$ with $\ell = n\lambda$ held fixed (see \cite{Lin:2022rbf, Rabinovici:2023yex} for how JT gravity emerges from the discrete theory in the case of the DSSYK model). 

\item The discrete phase space formalism seems to work best for prime $D$, and it would be nice to have a more uniform treatment for any $D$. Presumably one can get around this for non-prime dimensions by extending the Hilbert space to include some auxiliary states so as to enlarge the Hilbert space dimension to the nearest prime number. It would also be useful to generalize our results to the continuum setting. 

\item The notion of ``classicality'' that is familiar from AdS/CFT is one of large-$N$ factorization, i.e, large $N$ suppresses quantum fluctuations in some simple observables \cite{Jaffe}. On the other hand, Wigner negativity seems like it has more to do with efficiency of sampling the outcome probability distribution of a quantum circuit on a classical computer. It would be good to clarify how these two notions of classicality are related. 

\item The classical algorithms which seek to sample from the output probability distribution of the quantum circuit \cite{Mari_2012, Veitch_2012, Pashayan_2015, Wang_2019} are efficient when the Wigner quasiprobability distributions (or more precisely, their absolute values) for each of the circuit elements can be efficiently sampled. In a context where the circuit is made out of multiple qubits with local gates acting upon them, this is guaranteed by locality. In our context, one does not obviously have such a locality structure and so it is not straightforward to establish classical efficiency. It is worth noting that positive Wigner functions are highly constrained, and thus very special probability distributions. It is plausible that there is a sense in which they are easy to sample from, and that a small amount of negativity does not add much complexity. We would like to emphasize that this is an important issue to resolve in order to convincingly establish the connection between Wigner negativity and classicality in our context. 

\item In our discussion, it was important that the Hamiltonian be chaotic, to the extent that time evolution is ergodic; this makes it possible for the Krylov basis to span the entire Hilbert space. However, it would be interesting to understand the role of chaos v. integrability better. It would seem that in a free theory, the original description in terms of the elementary fields of the theory is already classical (at least for the time evolution of Gaussian states), so from a computational standpoint, there is no reason to look for a more efficient description. 

\item There is more to AdS/CFT than just Hamiltonian evolution -- one also needs to be able to interpret operator insertions in the boundary and the corresponding matter fields in the bulk within this framework. When Hamiltonian evolution is ergodic and spans the entire Hilbert space, the dual description presumably involves pure gravity. However, when this fails, we need additional operators (other than the Hamiltonian) in order to construct a basis for the Hilbert space. This should clearly result in a richer structure involving bulk matter.

\item There is a beautiful picture for the emergence of spacetime and gravity in holography from entanglement in the dual CFT \cite{VanRaamsdonk:2010pw, Lashkari:2013koa, Faulkner:2013ica, Faulkner:2017tkh, Lewkowycz:2018sgn} based on the Ryu-Takayanagi formula \cite{Ryu:2006bv}. It would be nice to understand how that picture fits with the complexity-based picture advocated here.

\item Finally, this discussion seems very tailored to $0+1$-dimensional quantum systems, and it would be interesting to explore whether gravity in higher dimensions could emerge from similar considerations.  
\end{enumerate}

\subsection{Other future directions}

\begin{figure}[t]
    \centering
    \includegraphics[height=6.6cm,width=0.85\linewidth]{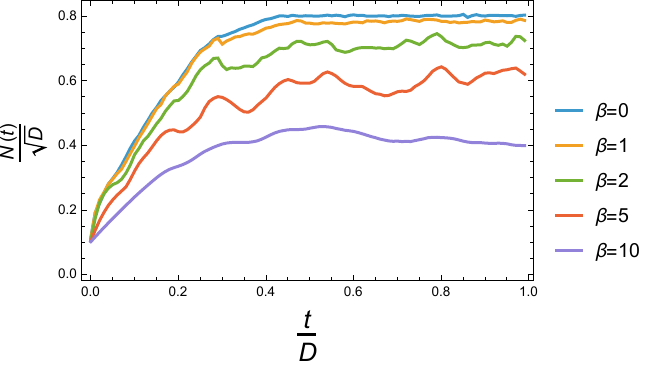}
    \caption{Temperature dependence of the negativity of the KW function for $D=101$.}
    \label{fig:KryTemp}
\end{figure}

\begin{figure}[t]
    \centering
     \includegraphics[height=6.6cm,width=0.85\linewidth]{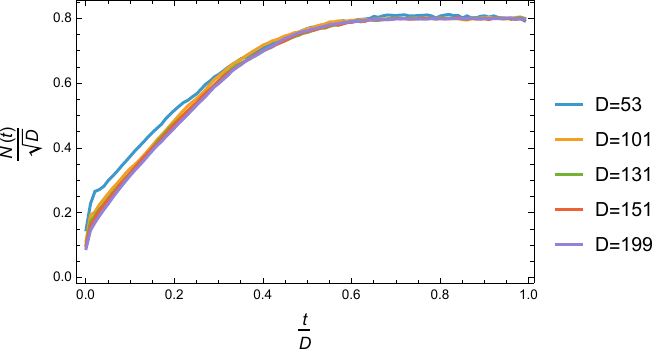}
    \caption{Wigner negativity for the infinite temperature thermofield double state for various values of $D$.}
    \label{fig:KryTFD}
\end{figure}

It would be useful to relate the Wigner negativity to the other notions of computational complexity developed in \cite{nielsen2005geometric, Nielsen_2006, Jefferson_2017, Chapman_2018, Brown_2018, Balasubramanian:2019wgd, Balasubramanian:2021mxo}. The quantity of interest in these works was the geodesic distance between an initial state/unitary and a final state/unitary with respect to a certain choice of complexity metric. The metric is such that motion along local directions is taken to be ``cheap'', while motion along non-local directions is penalized with a large cost factor. The Wigner negativity is not quite the same as circuit complexity, but one essential similarity is the classification of Clifford operations as being cheap and non-Clifford operations as being costly. One could imagine constructing a model of circuit complexity where one regards Clifford directions as being cost-free, and the remaining directions to have unit cost. It would be interesting to explore such notions of complexity and their connection with Wigner negativity and other magic monotones. 

In this paper, we have focused on the time-dependence of the KW-negativity in random matrix theory with the initial state being a generic pure state at infinite temperature. We can include finite temperature by taking the initial state to be of the form:
\beq 
|\psi_0\rangle= \frac{1}{\sqrt{Z_{\beta}}}e^{-\frac{\beta}{2}H}|0\rangle,
\eeq 
where $\beta$ is the inverse temperature. For inverse temperature $\beta \sim 1$, the time-dependence of the negativity is not affected very much (see figure \ref{fig:KryTemp} for some preliminary results). In particular, it shows the same general behavior, with a very similar saturation value. However, as the temperature is lowered, the saturation value of the negativity decreases (although it remains proportional to $\sqrt{D}$). It would be interesting to explore the temperature dependence of the negativity in more detail. It would also be interesting to consider other initial states, such as for example the thermofield double state (see figure \ref{fig:KryTFD} for some preliminary results in the infinite temperature case). Finally, it would be very interesting to explore these ideas in toy models for holography such as the SYK or the DSSYK model.

\subsection*{Acknowledgements}
We thank Sujay Ashok, Vijay Balasubramanian, Pawel Caputa, Jackson Fliss, Abhijit Gadde, Prahladh Harsha, Rohit Kalloor, Gautam Mandal, Shiraz Minwalla, Harshit Rajgadia and Sandip Trivedi for useful discussions and comments on an earlier version of the draft. We are grateful to Pruthvi Suryadevara for his significant help with numerical computations. We acknowledge supported from the Department of Atomic Energy, Government of India, under project identification number RTI 4002. We are grateful to the long term workshop YITP-T-23-01 held at YITP, Kyoto University, where part of this work was done. 

\appendix
\section{Early time approximation for the Wigner function}\label{sec:app}
In order to evaluate the KW function:
\beq \label{KWan}
W(Q,P) = \frac{1}{D}\sum_{k,\ell}\widehat{\delta}_{2Q,k+\ell}e^{\frac{2\pi i}{D}(k-\ell)P}\langle k|e^{-itH}|0\rangle \langle 0| e^{itH}|\ell\rangle,
\eeq
we need to evaluate the matrix elements $\langle k|e^{-itH}|0\rangle$. For $n \ll D$, the Krylov coefficients in GUE are approximately given by $b_n \sim 1$ and $a_n \sim 0$ \cite{Balasubramanian:2022tpr}. With this approximation, the Hamiltonian acting on the Krylov basis gives:
\beq 
H|n\rangle = |n+1\rangle + |n-1\rangle, \;\;\; \cdots\;\;\;(n \ll D).
\eeq 
So, for $Q \ll \frac{D-1}{2}$ and small times, we can evaluate the matrix elements in equation \eqref{KWan} by treating the system as a simple harmonic oscillator. Consider the operators
\beq 
E_+ = \sum_{n=0}^{\infty}|n+1\rangle\langle n|, \;\;E_- = \sum_{n=1}^{\infty} |n-1\rangle \langle n|.
\eeq 
in the simple harmonic oscillator. Then, the action of our Hamiltonian on states with $n \ll D$ takes the form 
\beq 
H \sim E_+ + E_-.
\eeq 

This description is, of course, approximate. Firstly, when $n$ starts scaling with $D$, the Krylov coefficients deviate significantly from 1. Secondly, the fact that the Hilbert space has $D$ states and is not infinite dimensional (as in the harmonic oscillator) becomes important. Nevertheless, we expect the approximation to be good for small times. In the harmonic oscillator, the eigenstates of the operator $(E_+ + E_-)$ are known \cite{SusskindGlogower, phase} (see also Appendix E of \cite{Rabinovici:2023yex}):
\beq 
|\theta\rangle = \sqrt{\frac{2}{\pi}}\sum_{n=0}^{\infty}\sin[(n+1)\theta]|n\rangle,
\eeq
\beq 
(E_+ + E_-) |\theta\rangle =2\cos\,\theta |\theta\rangle.
\eeq 
These states satisfy the completeness relation:
\beq 
\int_0^{\pi}d\theta\,|\theta\rangle \langle \theta| = 1.
\eeq 

\begin{figure}[t]
\centering
\includegraphics[height=6.6cm,width=10cm]{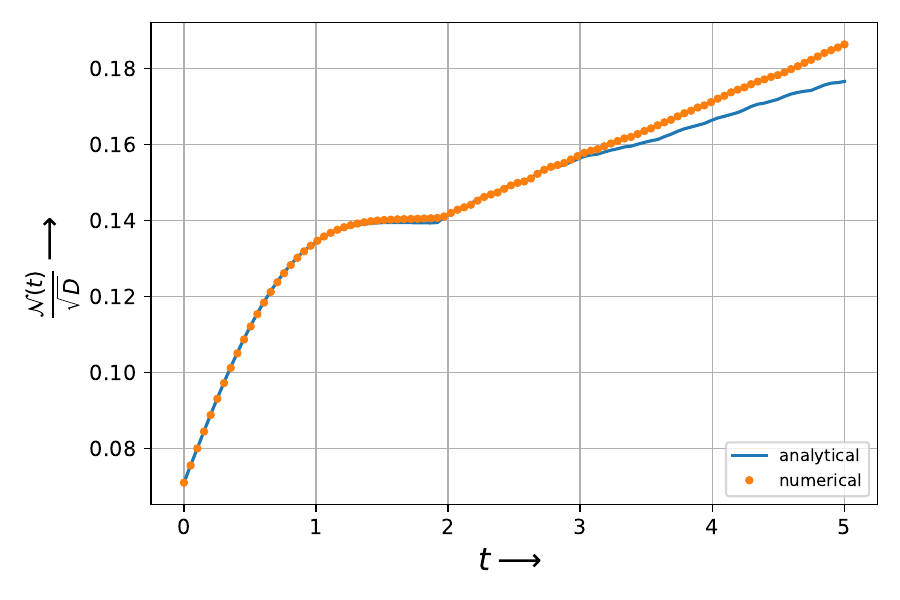}
\caption{A comparison of the Wigner negativity computed from the approximate formula \eqref{analytic} and the numerical result for D = 199.}
\label{fig:slope}
\end{figure}

We can use these relations to approximately evaluate the matrix elements of interest to us:
\beq
\langle k|e^{-itH}|0\rangle \sim \int_0^{\pi} d\theta\,\langle k|\theta\rangle\langle\theta |e^{-itH}|0\rangle  = \frac{2}{\pi}\int_0^{\pi} d\theta\,e^{-2it \cos\theta}\sin[(k+1)\theta]\sin(\theta).
\eeq
Thus, the Wigner function in this approximation is given by
\beq
W(Q,P) \sim \frac{4}{D \pi^2}\sum_{k,\ell}\delta_{2Q,k+\ell}e^{\frac{2\pi i}{D}(k-\ell)P} f_k(-t) f_{\ell}(t)
\eeq 
where 
\beqn 
f_k(t) &=& \int_0^{\pi}d\theta\,\sin\theta \sin[(k+1)\theta] e^{2it\cos \theta}\nonumber\\
&=&-\frac{1}{2}\int_0^{\pi}d\theta\,\left(\cos[(k+2)\theta] - \cos[k\theta]\right) e^{2it\cos \theta}\nonumber \\
&=& \frac{\pi i^k}{2}\left(J_{k+2}(2t) + J_k(2t)\right).
\eeqn 
Thus,
\beq \label{analytic}
W(Q,P) \sim \frac{i^{2Q}}{D}\sum_{k,\ell}\widehat{\delta}_{2Q,k+\ell}e^{\frac{2\pi i}{D}(k-\ell)P}\left(J_{k+2}(-2t)+J_{k}(-2t)\right)\left(J_{\ell+2}(2t)+J_{\ell}(2t)\right).
\eeq
This is a universal formula for the early time behaviour of the KW function in GUE. The negativity computed from this formula agrees well with the numerical computation at early times (see  figure \ref{fig:slope}).

\bibliographystyle{JHEP}
\bibliography{Reference_wigner.bib}
\end{document}